\batchmode
\makeatletter
\documentclass{article}
\usepackage{graphics}

\providecommand{\LyX}{L\kern-.1667em\lower.25em\hbox{Y}\kern-.125emX\@}

\newcommand{\lyxaddress}[1]{
  \par {\raggedright #1 
  \vspace{1.4em}
  \noindent\par}
}

\begin{document}

%%%%% USER-DEFINED MACROS HERE %%%%%

\def\RR{\hbox{{\rm I}\kern-.2em\hbox{\rm R}}}
\def\pRR{\hbox{{\tiny \rm I}\kern-.1em\hbox{{\tiny \rm R}}}}
\def\negro{\hspace*{\fill}$\blacksquare$}
\def\div{\mbox{{\rm div}}}
\def\rot{\mbox{{\rm rot}}}
\def\NN{\hbox{I\kern-.2em\hbox{N}}}
\def\eg{{e.g.\ }}
\def \l{\lambda}
%%%%%%%%%%%%%%%%%%%%%%%%%%%%%%%%%%%%%%%

%\hyphenation{super-lat-tice semi-con-ductor}

%\tighten       %Gives single-space (RevTex 3.0)

%\begin{document}
%\draft %prints PACS numbers in

\title{Kinetic theory of nucleation and coarsening}
%\renewcommand{\thesection}{\Roman{section}}
%\renewcommand{\thesubsection}{\Roman{section}.\Alph{subsection}}
%\renewcommand{\theequation}{\arabic{section}.\arabic{equation}}

%\date{\today %May 15, 2000}
\maketitle

\author{ J. C. Neu}
\lyxaddress{Department of Mathematics, 
University of California at Berkeley, Berkeley, 
CA 94720, USA}

\author{ L.L. Bonilla}
\lyxaddress{Escuela Polit\'{e}cnica Superior, 
Universidad Carlos III de Madrid, Avda.\
Universidad 30, 28911 Legan{\'e}s, Spain}

\begin{abstract}
Classical theory of nucleation based on Becker-Doering equations and coarsening for 
a binary alloy. 
\end{abstract}

%\pacs{PACS numbers: %05.45.+b, 05.20.-y, 05.40.+j
%}

%\begin{multicols}{2}
%\narrowtext

\setcounter{equation}{0}
\section{Introduction}
\label{sec-introduction}
The purpose of this chapter is to explain the classical kinetic
theory of nucleation in a context simpler than polymer
crystallization. Many theories start by assuming that polymer
crystallization is an activated process involving crossing of a
free energy barrier \cite{zia68}. The latter separates two
accessible stable states of the system such as monomer solution
and crystal. This general setting for activated processes can be
used to describe the formation of a crystal from a liquid
cooled below its freezing point \cite{mar95}, precipitation \index{precipitation} and
coarsening \index{coarsening} of binary alloys \cite{XH}, colloidal crystallization
\cite{gas01}, chemical reactions \cite{pag97}, polymer
crystallization \cite{zia68,reg01,zia01}, etc. In all these
cases, the theory of homogeneous isothermal nucleation provides
a framework to study the processes of formation of nucleii from
density fluctuations, and their growth until different nucleii
impinge upon each other. In the early stages of these
processes, nucleii of solid phase are formed and grow by
incorporating particles from the surrounding liquid phase.
There is a critical value for the radius of a nucleus that
depends on a chemical drive potential, which is proportional to
the supersaturation \index{supersaturation} for small values thereof. In this limit,
the critical radius \index{critical radius} is inversely proportional to the
supersaturation. At the beginning of the nucleation process,
nucleii have small critical radius and new clusters are being
created at a non-negligible rate. As the size of existing
clusters increases, there are less particles in the liquid
phase, the supersaturation decreases and the critical radius
increases. Then it is harder for new clusters to spontaneously
appear from density fluctuations. What happens is that
supercritical clusters (whose radii are larger than the
critical one) keep growing at the expense of subcritical
clusters, that in turn keep losing particles. The size of the
nucleii is still small compared to the average distance between
them, so that impingement processes (in which two or more
clusters touch and interaction between them dominates their
growth) can be ignored. This stage of free deterministic growth
is called {\em coarsening} \cite{LS}. 

A convenient framework to describe nucleation and coarsening is
the classical Becker-D\"oring \index{Becker-D\"oring} kinetic theory. We assume that the
dominant processes for nucleus growth or shrinking are addition
or subtraction of one particle. Nucleation is thus treated as a
chain reaction whereby nucleii of $n$ particles are created by
adding one particle to a nucleus of $n-1$ particles, or 
subtracting one particle from a nucleus with $n+1$ particles.
We can then write rate equations for the number density of
nucleii of $n$ particles by using the law of mass action. The
kinetic rate constants for the processes of addition and
depletion have to be determined by using specific information
from the physical process we are trying to model. Typically we
impose detailed balance which implies that the ratio of rate
constants is proportional to the exponential of the free energy
cost of adding one particle to a nucleus of $n$ particles (in
units of $k_B T$). This leaves one undetermined rate constant.
There are different ways of finding the missing constant. One
way is to postulate a microscopic theory for particle
interaction and use Statistical Mechanics to determine the free
energy of a cluster \cite{leb77}. A different point of view is
to impose that our rate constants should provide a description
of coarsening compatible with the macroscopic description in
terms of balance equations. We shall illustrate this second
point of view and be led to a Smoluchowski \index{Smoluchowski} equation from which
the Lifshitz-Slyozov \index{Lifshitz-Slyozov} coarsening theory follows \cite{juanjo}.

The structure of this paper is as follows. We present the
Becker-D\"oring kinetic equations for cluster with $n$
particles in Section \ref{sec-kclusters}. One relation between
the two rate constants of this theory follows from detailed
balance. The other rate constant has to be determined by
comparison with the known macroscopic equation for the growth of
cluster radii. In the small supersaturation limit, the
Becker-D\"oring equations can be approximated by a Smoluchowski
equation for the distribution function of cluster radii. Its
drift term yields the growth of cluster radii in terms of the
missing rate constant. To compare with experimental data, we
consider the case of coarsening of a binary alloy \cite{XH}. In
Sections \ref{sec-phaseeq} to \ref{sec-quasistatic}, we review
phase equilibria, macroscopic kinetics of precipitate and
matrix atoms and the quasistatic \index{quasistatic} limit of the kinetic
equations, respectively. As a result, we find the growth of the
radius of a supercritical cluster in terms of macroscopic
parameters. Comparison with the results in Section
\ref{sec-kclusters} yields the sought rate constant; see
Section \ref{sec-quasistatic}. Numerical values for all the
parameters involved in our theories can be calculated from
experimental data as explained in Section \ref{sec-XH}. A
discussion of our results constitutes the last Section.

\setcounter{equation}{0}
\section{Kinetics of clusters}
\label{sec-kclusters}
Let us assume that we have two stable phases
characterized by different values, $c_1$ and $c_2$ of the
number density $c$. Phase 1 is solution and Phase 2
precipitate. Or Phase 1 is the liquid and Phase 2 the
crystal phase. Initially all precipitate particles are in Phase
1. Classic Becker-D\"oring (BD) kinetics treats nucleation as a
{\em chain reaction} whereby nucleii (assumed to be {\em
spherical}) of
$n$ precipitate particles are created by adding one particle to
a nucleus with $n-1$ particles, or subtracting one particle
from a nucleus with $n+1$ particles. This chain reaction
scheme is natural for the situation that BD has
in mind, in which bulk precipitate phase consists
only of precipitate particles so distinction
between particles in nucleus or in solution is
clear. 

Let $\rho_n$ be the number density of nucleii of
$n$ particles. The monomer density $\rho_1$ represents
the concentration of precipitate particles in
solution and as such it will be identified with the
concentration $c_\infty$ of the macroscopic theory in Section
\ref{sec-quasistatic}. Consider the reaction
$$ n+1 \rightleftharpoons (n+1) .$$
The forward reaction proceeds at a rate proportional
to $\rho_1\,\rho_n$ with some rate constant $k_a$.
The backward reaction proceeds at a rate proportional
to $\rho_{n+1}$ with rate constant $k_d$. Hence the
net rate of creation of $(n+1)$-clusters from $n$-clusters per
unit volume is the {\em flux} 
\begin{eqnarray}
j_{n}\equiv k_{a,n}\rho_1\rho_n - k_{d,n+1}
\rho_{n+1}.   \label{5.1}
\end{eqnarray}
The fact that the rate constants depend on cluster
size has been explicitly indicated in (\ref{5.1}).
Net rate of creation of $n$-clusters is due to their creation
from $(n-1)$-clusters minus the rate of creation of
$(n+1)$-clusters from $n$-clusters, 
\begin{eqnarray}
\dot{\rho}_{n} = j_{n-1}- j_n\equiv - D_-\, j_n,
\quad n\geq 2.    \label{5.2}
\end{eqnarray}
This formulation specifies the evolutions of
$\rho_2$, $\rho_3$, \ldots with $\rho_1 = c_\infty$
{\em given}. The number density of precipitate
particles in $n$-clusters is $n\rho_n$ and the density
$c$ (equal to the initial concentration of precipitate particles
in the solution) of {\em all} precipitate particles is 
\begin{eqnarray}
c \equiv \sum_{n=1}^{\infty} n\,\rho_{n},   
\label{5.3}
\end{eqnarray}
or equivalently, 
\begin{eqnarray}
c - c_1 = \gamma_{\infty} + \sum_{n=2}^{\infty}
n\,\rho_{n}, \label{5.3b}
\end{eqnarray}
where we have defined $\gamma = c -c_1$, $\gamma_{\infty} =
c_{\infty} -c_1$. There is conservation of all precipitate
particles so $c$ or $\gamma_0 = c-c_1$ are constant.
(\ref{5.3b}) establishes that the constant initial
concentration disturbance
$\gamma_0$ is sum of the disturbance of precipitate particles,
$\gamma_\infty$, and the number density of particles in all
cluster sizes $n\geq 2$. Given the constraint (\ref{5.3}), the
evolution of all $\rho_n$, including $\rho_1$, is
specified. 

A most essential point of BD kinetics is
identification of rate constants $k_a$ and $k_d$.
Ideally, this would be based on basic energetics
and dynamics at the microscopic level, but a
complete realization of this ideal is clearly elusive. Here
is what {\em is} done: {\em The ratio is
determined by detailed balance}. Equilibrium, if
achievable, is described by a zero flux. Setting
$j_n =0$ in (\ref{5.1}) implies
\begin{eqnarray}
{k_{a,n}\over k_{d,n+1}} = {\rho_{n+1}\over\rho_{1}
\rho_{n}}\,.      \label{5.4}
\end{eqnarray}
Standard equilibrium physicochemical theory
states that  
\begin{eqnarray}
{\rho_{n+1}\over\rho_{n}} = e^{-{\mu_{n}\over
\tau}}\,,    \label{5.5}
\end{eqnarray}
where $\mu_n$ is the free energy cost of creating
an $(n+1)$-particle nucleus from an $n$-particle
nucleus relative to the state of no nucleus. $\tau= k_B T$ is
the temperature measured in units of energy. Clearly, $\mu_n =
G_{n+1} - G_n$ ($G_n$ is the free energy \index{free energy} of a
$n$-cluster), so (\ref{5.5}) becomes 
\begin{eqnarray}
{\rho_{n+1}\over\rho_{n}} = e^{-{G_{n+1} -
G_{n}\over\tau}}\,,    \nonumber
\end{eqnarray}
and (\ref{5.4}) now reads
\begin{eqnarray}
\rho_{1} k_{a,n} = e^{-{G_{n+1} -
G_{n}\over\tau}}\, k_{d,n+1}.    \nonumber
\end{eqnarray}
Thus formula (\ref{5.1}) for the flux is now
\begin{eqnarray}
j_{n} = k_{d,n+1}\,\left\{ e^{-{G_{n+1} -
G_{n}\over\tau}}\,\rho_n - \rho_{n+1}\right\} .   
\label{5.6}
\end{eqnarray}

The equilibrium considered in the detailed
balance argument is achievable only if $G_n\to
+\infty$ as $n\to\infty$, whereas in a
supersaturated solution, $G_n$ achieves a maximum
for finite $n$ and then $G_n\to -\infty$ as $n\to
\infty$. The determination of the ratio $k_a/k_d$
is assumed to hold regardless.

What is known about $G_n$? Microscopic models for
$G_n$ (or, equivalently, the cluster partition
function $Q_n \equiv \sum_K e^{U\, b(K)/\tau} =
e^{-G_{n}/\tau}$, where $U$ is the binding energy
per pair of particles in the cluster of $n$
particles, $b(K)$ is the number of
nearest-neighbor pairs of particles in the cluster
$K$, and the sum is over all translationally
inequivalent $n$-particle clusters) are described
in \cite{leb77,pen83,pen84}. We would like to
follow here a simpler approach, consisting of
identifying the resulting expressions for large
spherical nucleii with known facts about radius growth in
the quasistatic approximation. For nucleii of
macroscopic size $n\gg 1$, $n$ can be written in terms of the
cluster radius $a$ by
\begin{eqnarray}
n ={4\pi\over 3}\, c_2\, a^3,    \label{5.7}
\end{eqnarray}
and $G_n\sim G(a)$, where $G(a)$ is the free
energy of a nucleus of radius $a$ as determined
by continuum theory. For nucleii of only a few
particles, this asymptotic correspondence with
continuum theory breaks down. But if the critical
nucleus \index{critical nucleus} has $n\gg 1$ particles, the continuum
approximation works for $n$ on the order of
critical cluster size.

In the limit $|(G_{n+1}-G_n)/\tau| \ll 1$, formula (\ref{5.6})
for the flux reduces to 
\begin{eqnarray}
j_{n} = k_{d,n+1}\,\left\{ -{(G_{n+1} -
G_{n})\,\rho_{n}\over\tau} + \rho_n - \rho_{n+1}
\right\} \nonumber\\
= -  k_{d,n+1}\,\left\{ {1\over\tau}\, (D_+G_{n})
\,\rho_{n} + D_+ \rho_n \right\},   \label{5.8a}
\end{eqnarray}
where $D_{\pm} h_n \equiv \pm\, (h_{n\pm 1} -
h_n)$. The basic evolution equation (\ref{5.2})
now reads 
\begin{eqnarray}
\dot{\rho}_{n} - D_- \,\left\{ k_{d,n+1}\,\left(
{1\over\tau}\, (D_+ G_{n})\,\rho_{n} + D_+ \rho_n
\right)\right\} = 0.   \label{5.8b}
\end{eqnarray}
This equation looks like a spatially discretized
Smoluchowski equation. Asymptotic replacement of
difference operators $D_+$, $D_-$ by derivatives
is justified if the relative changes in $G_n$,
$\rho_n$ and $k_d$ when $n$ increases by one are
small. Here we will follow the simple procedure of
formulating the continuum limit an checking its validity a
posteriori. 

The space-like variable in (\ref{5.8b}) is $n$.
Experimental data usually contain histograms showing the
distribution of nucleii in the space of their radii $a$, so
we adopt the radius $a$ as a more convenient
space-like variable. The dependent variable
should be $\rho = \rho(a,t)$, the distribution of
nucleii in space of radius $a$. Thus $\rho(a,t)\,
da$ is the number of nucleii per unit volume with
radii in $(a,a+da)$. Conversions $n\to a$, $\rho_n
\to\rho$ are now determined. From (\ref{5.7}), it
follows that the change $da$ in $a$ when $n$
increases by 1 is given by
\begin{eqnarray}
1 \sim 4\pi\, c_2 a^2\, da.   \label{5.9}
\end{eqnarray}
In the general continuum theory of nucleii,
the concentration of precipitate particles inside a nucleus,
$c_2$, is a function of the radius $a$. But for many
experiments, deviations of $c_2$ from its equilibrium value for
a planar interface are negligible, so in (\ref{5.9})
any term arising from $a$-dependence of $c_2$ is dropped.
$\rho_n$ is related to $\rho(a,t)$ by 
\begin{eqnarray}
\rho_n \sim \rho(a,t)\, da\sim {1\over 4\pi
c_{2}}\, {\rho\over a^{2}}\,.  \label{5.10}
\end{eqnarray}
Given any sequence $h_n$ with continuum
approximation $h(a)$, 
\begin{eqnarray}
D_+ h_n \sim D_- h_n\sim {h_{a}\over 4\pi
c_{2}\, a^{2}}\,.  \label{5.11}
\end{eqnarray}
It follows from (\ref{5.10}) and (\ref{5.11}) 
that the continuum limit of (\ref{5.8b}) is 
\begin{eqnarray}
\rho_{t} - {\partial\over\partial a} \left\{ {k_{d}\over
(4\pi c_{2}a^{2})^{2}}\, \left( {\rho\over\tau}\,
{\partial G\over\partial a} + a^2 {\partial \over\partial a}
\left( {\rho\over a^{2}}\right)\right) \right\} = 0.
\label{5.12}
\end{eqnarray}
(Here $k_d$ is a function of $a$, to be specified). The constraint (\ref{5.3b}) can be
written as
\begin{eqnarray}
\gamma_\infty + c_2\, \int_0^\infty {4\pi\over
3}\, a^3\, \rho(a,t)\, da = c-c_1\equiv \gamma_0. 
\label{5.3c}
\end{eqnarray}

As time elapses, it will be seen that the diffusive term in Eq.\
(\ref{5.12}) becomes negligible in comparison with the drift
term. The latter yields the following equation for radius
growth: 
\begin{eqnarray}
\dot{a} = - {k_{d}\over (4\pi c_{2}a^{2})^{2}\tau}\,
{\partial G\over\partial a}\,. \label{dot-a}
\end{eqnarray}
We now present a macroscopic theory that gives an explicit
expression for $\dot{a}$ that can be compared to experimental
data. Then $k_d(a)$ can be determined and this will specify the
limit of $k_{d,n}$ for large $n$. 

\section{Phase equilibria \index{phase equilibria} of a binary material} 
\label{sec-phaseeq}

\subsection{Phase equilibria of a binary
material} 

Let us consider a medium consisting of two
different particles. The more abundant type is
called ``matrix'' \index{matrix}, the other ``precipitate'' \index{precipitate}.
Suppose that we have a uniform mixture at fixed
temperature and pressure. Let $\mu$ be the
chemical potential \index{chemical potential}, i.e., the free energy cost of
adding one precipitate particle to a pre-existing
solution. It is a function of the number density
of precipitate particles, $c$:
\begin{equation}
\mu =Ê\mu(c).\label{1.1}
\end{equation}
Let us now derive the relationship of the
chemical potential, $\mu(c)$, to the bulk free
energy density, $g(c)$. We shall add one
precipitate particle to a solution of total
volume $V$. Then the free energy changes from
$g(c) V$ to $g(c) V + \mu(c)$, but the volume
changes from $V$ to $V' \equiv V +\nu(c)$, where
$\nu(c)$ is the specific volume of a precipitate
particle in a solution of number density $c$. The
number density changes from $c$ to $c' \equiv
(cV+1)/(V+\nu)$, and therefore the new free
energy is also expressed as $g(c')V'$. Hence we
get the identity
\begin{equation}
g(c)V +Ê\mu(c) = g\left({cV+1\over V+\nu}
\right)\, (V+\nu).\label{1.2a}
\end{equation}
Since $cV\gg 1$, $\nu\ll V$, this identity
reduces to 
\begin{equation}
\mu(c) = g'(c) + \nu(c)\, \{g(c) - c g'(c)\}.
\label{1.2b}
\end{equation}
One can just as easily consider the chemical
potential $\overline{\mu}(c)$, which is the free
energy cost of adding one matrix particle to the
solution. Adding one matrix particle changes the
free energy to $g(c)V + \overline{\mu}(c)$. The
volume of the solution changes now to $V'=V +
\overline{\nu}(c)$, where $\overline{\nu}(c)$ is
the specific volume of a matrix particle in
solution. The concentration $c$ of the
precipitate changes to $c' \equiv cV/(V+
\overline{\nu})$, and therefore the new free
energy is $g(c')V'$. Hence we get an identity
analogous to (\ref{1.2a}),
\begin{equation}
g(c)V +Ê\overline{\mu}(c) = g\left({cV\over
V+\overline{\nu}} \right)\,
(V+\overline{\nu}). \label{1.3a}
\end{equation}
Again the conditions $cV\gg 1$, $\overline{\nu}
\ll V$ lead to an asymptotic reduction of
(\ref{1.3a}),
\begin{equation}
\overline{\mu}(c) = \overline{\nu}(c)\,\{g(c) - c
g'(c)\}.   \label{1.3b}
\end{equation}

\subsection{Phase equilibria in a dilute solution}
\begin{figure}
\begin{center}
{\par\centering \resizebox*{0.6\columnwidth}{!}{\includegraphics{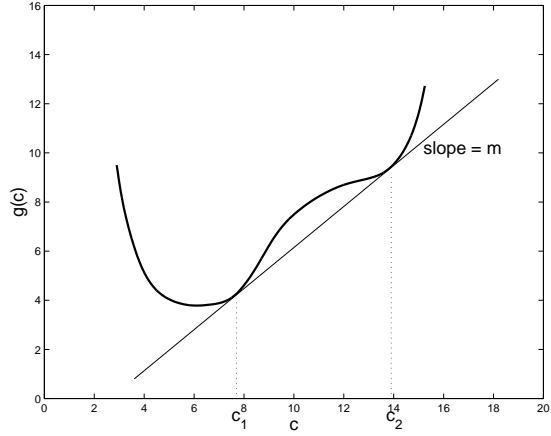}} \par}
%\vspace{0.5 cm}
\caption{Geometrical construction of stable phase
equilibria. }
\label{f1.1}
\end{center}
\end{figure}

Suppose that there are two stable phases
characterized by different values, $c_1$ and
$c_2$ of the number density $c$, and that these
phases occupy adjacent half spaces separated by a
planar interface. Imagine that one precipitate
particle is removed from phase 1 and dropped in
phase 2. The free energy cost is
$\mu(c_2) - \mu(c_1)$. In equilibrium, the energy
cost is to be zero, therefore
\begin{equation}
[\mu] \equiv \mu(c_2) - \mu(c_1) = 0. \label{1.4a}
\end{equation}
Similarly, the energy cost of removing a matrix
particle from phase 1 and dropping it in phase 2
is also to be zero, so
\begin{equation}
[\overline{\mu}]  = 0. \label{1.4b}
\end{equation}
The possible existence of multiple phases is
determined by the structure of $g(c)$. Consider a
{\em dilute} solution with $c$ much smaller than
the total atomic density. In this case, the
specific volumes $\nu$ and $\overline{\nu}$ of
precipitate and matrix particles should be nearly
constants independent of $c$: ``crowding'' effects
should be insignificant. In this case,
(\ref{1.3b}) and (\ref{1.4b}) imply
\begin{equation}
[g(c) - c \, g'(c)]\equiv [g-cg'] = 0.\label{1.5}
\end{equation}
Given this result, it now follows from
(\ref{1.2b}) and (\ref{1.4a}) that
$$[g'] = 0.$$
Hence,
\begin{equation}
g'(c_1) = g'(c_2) = M\quad \mbox{(common
value)},  \label{1.5a}
\end{equation}
and 
\begin{equation}
[g] - [c] \, M = 0.   \label{1.5b}
\end{equation}
Given $g(c)$ with $g''(c)<0$ in some interval of
$c$, and $g''(c)>0$ outside, one discerns the
standard geometrical construction of the solution
to (\ref{1.5a}) and (\ref{1.5b}) for $c_1$ and
$c_2$. This is depicted in Figure \ref{f1.1}.

\subsection{Critical nucleus}
Consider a spherical nucleus of phase 2
surrounded by phase 1. The energy cost of adding
one precipitate particle to this nucleus is
$$[\mu] + 8\pi\sigma r\, dr.$$
Here $r$ is the initial radius of the nucleus,
$dr$ is the change in radius due to adding one
precipitate particle, and $\sigma$ is the surface
tension. One has 
$$4\pi r^2 dr = \nu$$
so the energy cost can be written as 
$$[\mu] + {2\sigma\nu\over r}\,.$$
For a nucleus in equilibrium, this energy cost is
zero, therefore
\begin{equation}
[\mu] = - {2\sigma\nu\over r}\,.\label{1.6a}
\end{equation}
Now we add one matrix particle to the nucleus. No
energy cost for this process implies
\begin{equation}
[\overline{\mu}] = - {2\sigma\overline{\nu}\over
r}\,.\label{1.6b}
\end{equation}
After substituting for $\overline{\mu}$ from
(\ref{1.3b}), this equation becomes 
\begin{equation}
[g - c\, g'] = - {2\sigma\over r}\,. 
\label{1.7}
\end{equation}
Now substitute (\ref{1.2b}) for $\mu$ and
(\ref{1.7}) for $2\sigma/r$ in (\ref{1.6a}) to
get 
\begin{equation}
[g'] = 0,\label{1.8}
\end{equation}
which is the same as in the case of a planar
interface, (\ref{1.5a}). Given the concentration
$c_1$ of phase 1, this equation determines the
concentration $c_2$ inside the nucleus, and then
the Gibbs-Thomson \index{Gibbs-Thomson} relation (\ref{1.6a}) determines
the radius of the nucleus, $r$. This
determination simplifies when the concentrations
$c_1$ and $c_2$ are near ``planar'' values and
$\gamma_1$ and $\gamma_2$ are deviations from the
planar values. (\ref{1.8}) together with
(\ref{1.5}) for the planar case imply 
\begin{equation}
[g''\,\gamma] = 0.\label{1.9}
\end{equation}
Let us denote the common values of $g''(c_1)\,
\gamma_1$ and $g''(c_2)\, \gamma_2$ by $m$. The
variation of $[g-c\, g']$ in (\ref{1.7}) is 
\begin{equation}
- [c\, g''\,\gamma] = -[c]\, m.\nonumber
\end{equation}
Hence  (\ref{1.7}) gives 
\begin{equation}
[c]\, m = {2\sigma\over r}\,.\label{1.10}
\end{equation}

\section{Macroscopic kinetics}
\label{secMkinetics}
\subsection{Balance equations and jump conditions}
Let $c(x,t)$, $\overline{c}(x,t)$ denote the 
macroscopic number densities of precipitate and
matrix atoms. Local volume fractions of
precipitate and matrix atoms are $\nu c$ and
$\overline{\nu}\, \overline{c}$, respectively.
Since matrix and precipitate atoms fill space
leaving no gaps, we have the {\em space filling
condition}
\begin{equation}
\nu\, c + \overline{\nu}\, \overline{c} = 1. 
\label{2.1}
\end{equation}
In conventional kinetics, the density of
precipitate is locally conserved, with a flux
proportional to the gradient of the precipitate
chemical potential $\mu(c)$,
$$ c_t + \nabla\cdot (-\delta\,\nabla \mu) = 0,
$$ 
or
\begin{equation}
c_t =  \nabla\cdot (D\,\nabla c),\quad D =
\delta(c)\, \mu'(c). 
\label{2.2}
\end{equation}
Here $\delta(c)$ is a positive mobility
coefficient and $-\delta\nabla\mu = - D\,\nabla
c$ is the flux of $c$. This flux is formally a
diffusion with diffusion coefficient $D =
\delta(c)\,\mu'(c)$. Given $\mu(c)$ as in
(\ref{1.2b}), and provided $\nu$ and
$\overline{\nu}$ {\em do not depend on} $c$
(dilute solution), 
\begin{equation}
D = \delta(c)\,\mu'(c) = \delta (1-\nu c)\,
g''(c) = \delta\overline{\nu}\,\overline{c}\,
g''(c).      \label{2.3a}
\end{equation}
In the last equality, the space filling condition
has been used to replace $1-\nu c$ by
$\overline{\nu}\,\overline{c}$. For {\em stable}
bulk phases, $D$ must be positive, and
(\ref{2.3a}) then implies $g''(c)>0$. The
description of matrix transport is essentially
the same. The flux of matrix concentration
$\overline{c}$ is
\begin{equation}
-\overline{\delta}\,\nabla\overline{\mu} =
-\overline{\delta}\,\overline{\mu}'\, \nabla
\overline{c} = - \overline{D}\,\nabla\overline{c} 
\nonumber
\end{equation}
where $\overline{\mu}(\overline{c})$ is the
chemical potential of the matrix atoms as a
function of the matrix concentration
$\overline{c}$, $\overline{\delta}(c)$ is the
mobility of the matrix atoms, and $\overline{D}$
is the matrix diffusion coefficient given by 
$$
\overline{D} =\overline{\delta}\,\overline{\mu}'.
$$
Let $\overline{g}(\overline{c})$ be the free
energy density {\em as a function of} 
$\overline{c}$. The space filling condition and
$g(c) =\overline{g}(\overline{c})$ imply that
$$\overline{\mu}= \overline{g}'(\overline{c}) + 
\overline{\nu}\, \{\overline{g}(\overline{c}) -
\overline{c}\,\overline{g}'(\overline{c})\},$$ 
which is totally symmetric to (\ref{1.2b}). Then
the diffusion coefficient $\overline{D}$ is
related to $\overline{g}(\overline{c})$ by a
formula symmetric to (\ref{2.3a}), 
\begin{equation}
\overline{D} = \overline{\delta}\nu c
\overline{g}''(\overline{c}) . \label{2.3b}
\end{equation}
The space filling condition leads to a relation
between the mobilities $\delta$ and $\overline
\delta$. The linear combination $\nu c+\overline{
\nu}\overline{c}$ is locally conserved with flux
$-\nu D\nabla c - \overline{\nu} \overline{D}
\nabla \overline{c}$. But $\nu c+\overline{
\nu}\overline{c}\equiv 1$, so this flux is
divergence free. Let $C$ be any closed surface.
Use of divergence theorem yields 
$$\int_C (\nu\, Dc_n +\overline{\nu}\,\overline{D}
\overline{c}_n)\, da = 0.
$$
The space filling condition implies $\nu c_n = -
\overline{\nu}\overline{c}_n$, so that we get 
$\int_C \nu\, (D-\overline{D})\, c_n\, da = 0$.
This holds for all concentration fields $c$ and
closed surfaces $C$. Hence,
\begin{equation}
D=\overline{D} \label{2.4a}
\end{equation}
and by (\ref{2.3a}) and (\ref{2.3b}), 
\begin{equation}
\delta(c)\,\overline{\nu}\, \overline{c}\,
g''(c) = \overline{\delta}(\overline{c})\,\nu c\,
\overline{g}''(\overline{c}) . \label{2.4b}
\end{equation}
Now, 
$$g(c) = \overline{g}(\overline{c}) =
\overline{g}\left({1\over\overline{\nu}}\,
(1-c\nu)\right),$$
therefore, provided again that $\nu$ and
$\overline{\nu}$ {\em do not depend on} $c$,
$$g''(c) = \left({\nu\over\overline{\nu}}
\right)^2\, \overline{g}''(\overline{c}), $$
and (\ref{2.4b}) becomes
\begin{equation}
\delta\,\nu\,\overline{c} = \overline{\delta}\,
\overline{\nu}\, c . \label{2.5}
\end{equation}
This is the relation between mobilities. 

The integral form of Equation (\ref{2.2}) informs
the upcoming discussion of boundary conditions on
a phase interface. Let $R=R(t)$ be a time
sequence of closed regions in which $c=c(x,t)$ is
a smooth solution of (\ref{2.2}). The number of
precipitate particles in $R$ is 
$$N = \int_R c\, dx, $$
and the time rate of change is 
$$\dot{N} = \int_R c_t\, dx + \int_{\partial R} U
c\, da, $$
where $U$ is the outward normal velocity of
$\partial R$. Using (\ref{2.2}) to substitute for
$c_t$ above, and then the divergence theorem, we
obtain
$$\dot{N} = \int_{\partial R} (U\, c+ D\, c_n)\,
da.$$
The interpretation of this equation is clear:
The influx of precipitate atoms per unit area on
$\partial R$ is 
\begin{equation}
U\, c + D\, c_n . \label{2.6}
\end{equation}

Suppose now that there is a region $R_1$ of
matrix surrounding a region $R_2$ of precipitate.
The influx of precipitate atoms per unit area on
interface $C$ is given by (\ref{2.6}) with $c$,
$c_n$ evaluated on the precipitate side of $C$.
The outflux of precipitate atoms from the
surrounding matrix into $R_2$ is also given by
(\ref{2.6}) with $c$, $c_n$ evaluated on the {\em
matrix} side of $C$. Since precipitate atoms do
not accumulate on $C$ to form a surface density,
the following jump condition holds 
\begin{equation}
[U\, c + D\, c_n] = 0\quad\mbox{on $C$}.
\label{2.7}
\end{equation}

 In summary, conservation of precipitate is
expressed by the diffusion equation (\ref{2.2})
and the associated jump condition (\ref{2.7}).
The matrix density $\overline{c}$ satisfies the
diffusion equation and jump condition with the
same diffusion coefficient $D$. The space filling
condition (\ref{2.1}) is automatically upheld by
this kinetics. In addition there are
``thermodynamic'' jump conditions
\begin{eqnarray}
[g'] = 0, \label{2.8a}\\
\left. [ g - c g' ] = - 2\sigma\kappa, \right.
\label{2.8b}
\end{eqnarray}
expressing local equilibrium about the phase
interface. These are in fact Equations
(\ref{1.7}), (\ref{1.8}) with $1/r$ replaced by
the mean curvature $\kappa$. Equations
(\ref{2.2}), (\ref{2.7}), (\ref{2.8a}) and
(\ref{2.8b}) constitute a free boundary problem
for the evolution of precipitate concentration
$c$ and the phase interfaces. 

\subsection{Time evolution of Gibbs energy}
The total Gibbs free energy is 
\begin{eqnarray}
G = G_1 + G_2 + \sigma S,  \label{2.9a}
\end{eqnarray}
where 
\begin{eqnarray}
G_1 \equiv  \int_{R_{1}} g\, dx, \quad 
G_2\equiv  \int_{R_{2}} g\, dx \label{2.9b}
\end{eqnarray}
are free energies of bulk matrix and precipitate
phases, and $S$ is the surface area of the phase
interface, and $\sigma$ the surface tension. The
time evolution of $G$ under the kinetics of the
free boundary problem (\ref{2.2}), (\ref{2.7}),
(\ref{2.8a}) and (\ref{2.8b}) is examined here.
To compute the rates of change of $G_1$ and
$G_2$, it is useful to formulate a transport
equation for the free energy density $g(c)$: 
$$ g_t = g'\, c_t = g'\, \nabla\cdot (D\nabla c)
= \nabla\cdot (D\nabla g) - D\, g''\,|\nabla c|^2
$$
or
\begin{eqnarray}
g_t - \nabla\cdot (D\nabla g) = - D\, g''|\nabla
c|^2 . \label{2.10}
\end{eqnarray}

In stable bulk phases, $g''>0$, therefore the
source density in (\ref{2.10}) is generally
negative. Let us now look at the rate of $G_2$. 
\begin{eqnarray}
\dot{G}_2 = \int_{R_{2}} g_t\, dx + \int_C
U\, g\, da, \label{2.11}
\end{eqnarray}
where $U$ is the normal velocity of the phase
interface, positive if outward from precipitate. 
Inserting $g_t$ from (\ref{2.10}) and using the
divergence theorem, (\ref{2.11}) becomes 
\begin{eqnarray}
\dot{G}_2 = \int_C (U\, g+D\, g_n)\big|_2 \, da -
\int_{R_{2}} D g'' |\nabla c|^2 \, dx. 
\label{2.12a}
\end{eqnarray}
In the surface integral, the subscript 2 means
evaluation on precipitate side of interface.
Similarly, 
\begin{eqnarray}
\dot{G}_1 = -\int_C (U\, g+D\, g_n)\big|_1\, da -
\int_{R_{1}} D g'' |\nabla c|^2 \, dx,
\label{2.12b}
\end{eqnarray}
where subscript 1 means evaluation on matrix side
of interface. The rate of change of the surface
energy $\sigma S$ in (\ref{2.9a}) is given by the
standard formula of differential geometry, 
\begin{eqnarray}
\sigma\dot{S} = 2\sigma\int_C \kappa\, U\, da.
\label{2.12c}
\end{eqnarray}
Adding Equations (\ref{2.12a}) to (\ref{2.12c}),
we obtain the rate of change of the total free
energy
\begin{eqnarray}
\dot{G} = &-& \int_{R_{1}+R_{2}} D g'' |\nabla
c|^2\, dx \nonumber\\
&+& \int_C \{ [U\, g+D\, g_n] +
2\sigma\kappa U\}\, da,     \label{2.13}
\end{eqnarray}
Here the jump $[\dots ]$ denotes values on
precipitate side minus values on matrix side.
Using the continuity condition $[g']=0$, it
follows that
$$ [Ug + Dg_n] = U\, [g] + g'\, [D\, c_n].$$
In the right hand side, $g'$ denotes a well
defined value on the phase interface. By
conservation jump condition (\ref{2.7}), $[D\,
c_n] = - U\, [c]$, hence
\begin{eqnarray}
[Ug + Dg_n] = U \, [g-c g']. \label{2.14}
\end{eqnarray}
By the thermodynamic jump condition (\ref{2.8b}),
$[g - cg'] = -2\sigma\kappa$, so finally,
$$[Ug + Dg_n] = - 2\sigma U \kappa,
$$
and the energy rate formula (\ref{2.13}) reduces
to
\begin{eqnarray}
\dot{G} = -\int_{R_{1}+R_{2}} D g'' |\nabla c|^2
\, dx.     \label{2.15}
\end{eqnarray}
The integral is negative definite. Notice that
surface integral contributions to $\dot{G}$
cancel. This point is examined by direct physical
argument to see what it really means. 

Recall that influx of precipitate atoms into
precipitate phase per unit area is 
$$ U\, c + D\, c_n.$$
The free energy of each precipitate atom changes
by an amount $[\mu] = - 2\sigma\kappa\nu$,
according to (\ref{1.6a}), as it crosses from
matrix to precipitate. Hence there is a
contribution to $\dot{G}$ of 
\begin{eqnarray}
- 2\sigma\, (U\nu c + D\nu c_n)\, \kappa    
\label{2.16a}
\end{eqnarray}
per unit area of phase interface due to crossing
of precipitate atoms. Similarly, influx of matrix
atoms into precipitate phase per unit area is
$$ U\, \overline{c} + D\, \overline{c}_n$$
and change in free energy for each matrix atom
crossing into precipitate is $[\overline{\mu}] = -
2\sigma\kappa\overline{\nu}$, according to
(\ref{1.6b}). Hence, crossing of matrix atoms
gives another surface contribution to $\dot{G}$,
of
\begin{eqnarray}
- 2\sigma\, (U\overline{\nu}\, \overline{c} + D
\overline{\nu}\, \overline{c}_n)\,\kappa    
\label{2.16b}
\end{eqnarray}
per unit area. Adding (\ref{2.16a}) and 
(\ref{2.16b}) yields surface contribution to
$\dot{G}$ due to crossing of both types of atoms, 
$$- 2\sigma\, \{ U\, (\nu c +\overline{\nu}\,
\overline{c}) + D\, (\nu c_n + \overline{\nu}\,
\overline{c}_n)\}\,\kappa = - 2\sigma U\kappa
$$
per unit area. Here the space-filling constraint
has been used. From (\ref{2.12c}) it is seen that
$2\sigma U\kappa$ can be identified as a rate of
change of surface energy per unit area. Hence,
the total rate of free energy production per unit
area of phase interface is
$$- 2\sigma U\kappa + 2\sigma U\kappa = 0.
$$

\section{Quasistatic nuclei}
\label{sec-quasistatic}

An isolated region $R_2$ of precipitate phase,
called a {\em nucleus} is assumed spherical, and
concentration field $c$ is assumed spherically
symmetric. The kinetics is quasistatic if the
time derivative in the diffusion equation
(\ref{2.2}) is negligible. Here kinetics is
analyzed under the quasistatic assumption and
regimes of validity are determined a posteriori
by the criterion
\begin{eqnarray}
{Ua\over D} = {a\dot{a}\over D}\ll 1.   
\label{3.1}
\end{eqnarray}
Here $a$ is the radius of the nucleus, and the
normal velocity $U=\dot{a}$. $a^2/D$ represents
the characteristic time of diffusive transport in
the matrix phase surrounding the nucleus. The
characteristic time associated with the kinetics
of the radius is $a/U=a/\dot{a}$. Kinetics is
quasistatic if the time scale of the radius is
much longer than the diffusion time in the
surrounding matrix, as in (\ref{3.1}). 

\begin{figure}
\begin{center}
{\par\centering \resizebox*{0.6\columnwidth}{!}{\includegraphics{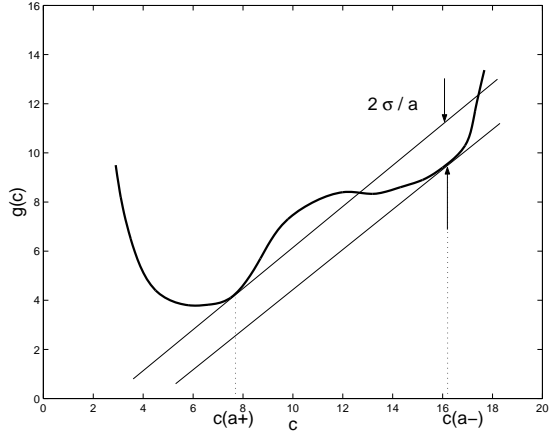}} \par}
%\fbox{
%\epsfig{%file=curva2.eps,width=7cm
%}}
\vspace{0.5 cm}
\caption{Geometrical construction of $c(a-)$ and
$c(a+)$. }
\label{f3.1}
\end{center}
\end{figure}

Under assumptions of radial symmetry and
quasistatic kinetics, the diffusion equation
(\ref{2.2}) reduces to
\begin{eqnarray}
\partial_r (r^{2}D c_{r}) = 0. \label{3.2}
\end{eqnarray}
The conservation jump condition (\ref{2.7}) reads
\begin{eqnarray}
\dot{a}\, [c] = - [Dc_{r}]. \label{3.3}
\end{eqnarray}
The thermodynamic jump conditions (\ref{2.8a})
and (\ref{2.8b}) read
\begin{eqnarray}
[g'] = 0, \label{3.4a}\\
\, [g-cg'] = - {2\sigma\over a}\,. \label{3.4b}
\end{eqnarray}
Given the value $c_\infty$ of $c$ as
$r\to\infty$, Equations (\ref{3.2}) to (\ref{3.4b}) determine
an ordinary differential equation (ODE) for the nuclear radius
$a(t)$. The first integral of (\ref{3.2}) is 
\begin{eqnarray}
r^{2}D c_{r} = Q, \quad\mbox{or}\quad D c_{r} =
{Q\over r^{2}}.   \label{3.5}
\end{eqnarray}
Here $Q$ is a function of time on $r<a$ or $r>a$,
but with a possible jump at $r=a$. Regularity of
$c$ at $r=0$ forces $Q\equiv 0$ on $r<a$. From
(\ref{3.5}) it is now evident that
\begin{eqnarray}
[D c_{r}] = - {Q\over a^{2}}\,,  \nonumber
\end{eqnarray}
where $Q$ now refers to the value on $r>a$. The
conservation jump condition (\ref{3.3}) now reads 
\begin{eqnarray}
\dot{a} [c] = {Q\over a^{2}}\,.   \label{3.6}
\end{eqnarray}
This is a differential equation for $a(t)$, once
the dependence of $[c]$ and $Q$ upon the radius
are determined. The two thermodynamic jump
conditions (\ref{3.4a}) and (\ref{3.4b})
determine $c(a-)$ and $c(a+)$, hence $[c]$ as a
function of $a$. Figure \ref{f3.1} shows the
graphical construction of (\ref{3.4a}) and
(\ref{3.4b}). 

To determine $Q$, integration of (\ref{3.5}) in
$r>a$ is required. Let $h(c)$ be a function such
that $h'(c) = D(c)$. (\ref{3.5}) now reads
\begin{eqnarray}
\partial_{r} h(c) = {Q\over r^{2}}\,,  \nonumber
\end{eqnarray}
and integration from $r=a$ to $r=\infty$ gives
\begin{eqnarray}
h(c_{\infty}) - h(c_+) = {Q\over a}\quad \mbox{or}
\quad Q = a\,\{h(c_{\infty}) - h(c_+)\}\,.
\nonumber
\end{eqnarray}
Here $c_+$ means $c(a+)$. (\ref{3.6}) now reads
\begin{eqnarray}
\dot{a} = {h(c_{\infty}) - h(c_{+})\over a\,
[c]}\,.   \label{3.7}
\end{eqnarray}
Since $c_+$ and $[c]$ are definite functions of
$a$ as determined by the thermodynamic jump
conditions (\ref{3.4a}) and (\ref{3.4b}),
(\ref{3.7}) is the required ODE for $a(t)$.

\subsection{Small supersaturation}
A standard limit called {\em small
supersaturation} is realized when the
concentration is close to planar equilibrium
values $c=c_1$ in matrix phase $r>a$ and $c=c_2$
in precipitate phase $r<a$. Let $\gamma$ denote
disturbance of $c$ from planar equilibrium
values: $\gamma=c-c_1$ in matrix phase,
$\gamma=c-c_2$ in precipitate phase. For
$|\gamma| \ll c_1, c_2$, (\ref{3.7}) reduces to
\begin{eqnarray}
\dot{a} \sim {D_{1}\, (\gamma_{\infty} -
\gamma_{+})\over a\, [c]}\,.   \label{3.8}
\end{eqnarray}
Here $D_1\equiv D(c_1)$ and $[c] = c_2 - c_1$.
Also $\gamma_{\infty}=\gamma(r=\infty)$ and
$\gamma_+ = \gamma(a+)$. $\gamma_+$ is determined
from asymptotic limit of (\ref{3.4a}) and
(\ref{3.4b}): Reduction of (\ref{3.4a}) is 
\begin{eqnarray}
[g'' \gamma] =0\Longrightarrow g''(c_1) \gamma_+
= g''(c_2)\gamma_{-} = m\,\mbox{(common value),}  
\label{3.9a}
\end{eqnarray}
and given $m$, reduction of (\ref{3.4b}) is 
\begin{eqnarray}
[c]\, m = {2 \sigma\over a}\quad\mbox{or}\quad 
g''_1\,Ê[c]\,\gamma_{+} = {2 \sigma\over a}\,.
\label{3.9b}
\end{eqnarray}
Substituting this result for $\gamma_+$ into
(\ref{3.8}) gives the reduced ODE
\begin{eqnarray}
\dot{a} \sim {D_{1}\over [c]\, g''_{1}}\, {1\over
a}\,\left(g''_1\, \gamma_{\infty} - {2 \sigma
\over a\, [c]}\right)\,. \label{3.10}
\end{eqnarray}
This equation indicates that clusters whose radii are smaller
than the critical value
\begin{eqnarray}
a_c \equiv  {2\sigma\over g''_{1}\,\gamma_{\infty}
\, [c]} \,, \label{3.11}
\end{eqnarray}
shrink and disappear. Supercritical clusters of radius larger
than the critical radius $a=a_c$ grow steadily according to
(\ref{3.11}).

Is this small supersaturation kinetics consistent
with the quasistatic criterion (\ref{3.1})?
Natural unit of $a$ is $a_c$ given by (\ref{3.11}),
which is the standard formula for critical radius
in small supersaturation limit. Given $a=O(a_c)$,
an order of magnitude estimate of $\dot a$ based
on (\ref{3.10}) is 
\begin{eqnarray}
\dot{a} = O\left( {D_{1}\gamma_{\infty}\over
a_{c}\, [c]}\right)\nonumber
\end{eqnarray}
which can be rearranged as
\begin{eqnarray}
{a\dot{a}\over D_{1}} = O\left( {\gamma_{\infty} 
\over [c]}\right)\label{3.12}
\end{eqnarray}
The analysis here is based on $|\gamma|\ll c_1 ,
c_2$. But for quasistatic kinetics, we need 
\begin{eqnarray}
|\gamma_{\infty}|\ll [c]= c_2 - c_1.\label{3.13}
\end{eqnarray}

The reduced ODE (\ref{3.10}) indicates natural
units of $\gamma$, space and time. The limit
(\ref{3.13}) is embodied by measuring $\gamma
\equiv c-c_e$ in units of $\epsilon\, c_e$, where
$\epsilon>0$ is a gauge parameter and limit
$\epsilon\to 0+$ is considered. Given this unit
of $\gamma$, the order of magnitude of $a_c$ in
(\ref{3.11}) is $l/\epsilon$, $l\equiv
\sigma/([c] g''_1 c_e)$. $l/\epsilon$ is adopted
as unit of length. Finally, the unit of time
which gives $\dot a$ as in (\ref{3.12}) is
$\tau_e/\epsilon^3$, $\tau_e\equiv (l^2/D_1)\,
([c]/c_e)$. This system of units is summarized in
the following Table:\\

\begin{tabular}{||c|c||cr||}    
\hline Variable& Unit\\
\hline $\gamma\equiv c - c_e$ & $\epsilon c_e$  
\\
\hline $x$& ${l\over\epsilon}$, $l\equiv {\sigma
\over [c] g''_{1} c_{e}}$
\\
\hline $t$ & ${\tau_{e}\over\epsilon^{3}}$,
$\tau_e \equiv {l^{2} [c]\over D_{1} c_{e}}$\\
\hline $G$ & ${\sigma l^{2}\over \epsilon^{2}}$\\
\hline
\end{tabular}
\\
\\
{\em Scaling Table for small supersaturation.}
\bigskip

Given these units, the nondimensional version of
(\ref{3.10}) is 
\begin{eqnarray}
\dot{a} = {1\over a}\,\left(\gamma_{\infty} -
{2\over a}\right)\,.\label{3.14}
\end{eqnarray}

\subsection{Quasistatic energetics}
Given evolution of precipitate concentration
field $c$, corresponding changes in total free
energy of medium are quantified by the rate
formula (\ref{2.15}). Now suppose concentration
field $c$ corresponds to a nucleus undergoing
quasistatic evolution as set by criterion
(\ref{3.1}). It seems reasonable to approximate
the integral in (\ref{2.15}) using the
quasistatic approximation to $c$ which satisfies
Equations (\ref{3.2}) to (\ref{3.4b}). The rate
formula (\ref{2.15}) reduces to 
\begin{eqnarray}
\dot{G} = -\int_{0}^{\infty} D g'' c_r^2 4\pi r^2
\, dr.\nonumber
\end{eqnarray}
>From (\ref{3.5}), $r^2 D c_r =Q$ for $r>a$, and
$r^2 D c_r=0$ for $r<a$, so this reduces to
\begin{eqnarray}
\dot{G} = - 4\pi Q\int_{a+}^{\infty} g'' c_r\,
dr = - 4\pi Q\,\{ g'(c_{\infty}) - g'(c_+)\}.
\nonumber
\end{eqnarray}
Substituting for $Q$ from (\ref{3.6}), 
\begin{eqnarray}
\dot{G} = - [c]\,\{ g'(c_{\infty}) -
g'(c_+)\} (4\pi a^{2} \dot{a}). \label{3.15}
\end{eqnarray}
In the right hand side, $[c]$ and $g'(c_+)$ are
definite functions of radius $a$, determined by
thermodynamic jump conditions (\ref{3.4a}) and
(\ref{3.4b}). Hence, in quasistatic evolution,
$G$ is effectively a function of the radius $a$.
>From (\ref{3.15}), it is seen that $G(a)$ obeys
the differential relation
\begin{eqnarray}
dG = - [c]\,\{ g'(c_{\infty}) -
g'(c_+)\} (4\pi a^{2} da). \label{3.16}
\end{eqnarray}
In conventional descriptions of nucleus
energetics, the Gibbs free energy cost of a
nucleus regarded as a function of the instantaneous
radius $a$, independent of past history. The
specific formula is
\begin{eqnarray}
G = 4\pi\sigma a^2 - g_{b}{4\pi\over 3}\,
a^{3}  \label{3.17a}
\end{eqnarray}
or in differential form, 
\begin{eqnarray}
dG = 8\pi \sigma a da - g_b 4\pi a^{2} da.
\label{3.17b}
\end{eqnarray}
Here $\sigma$ is surface tension and $4\pi\sigma
a^2$ represents surface energy. $g_b$ is a
constant with units of energy density, sometimes
called ``chemical driving force''. It is the free
energy released per unit volume of increase of
precipitate phase. In its present form
(\ref{3.16}) does not have obvious correspondence
to (\ref{3.17b}). A correspondence is brought out
by reformulation of (\ref{3.16}) with help of
thermodynamic jump conditions (\ref{3.4a}) and
(\ref{3.4b}). By (\ref{3.4a}), it follows that
$g'(c_-) = g'(c_+)=$ common value $m$, and hence 
(\ref{3.4b}) gives 
$$[g] - [c]\, m = - {2\sigma\over a}\,,$$
or, equivalently,
$$[c]\, g'(c_+) = [g] + {2\sigma\over a}\,.$$
Hence (\ref{3.16}) becomes
\begin{eqnarray}
dG = 8\pi\sigma\, a\, da + \{ [g] - [c]\,
g'(c_{\infty})\} (4\pi a^{2} da). \label{3.18}
\end{eqnarray}
The first term on RHS is differential of surface
energy, same as in (\ref{3.17b}). An apparent
correspondence between (\ref{3.17b}) and
(\ref{3.18}) is completed by identifying the
chemical driving force $g_b$,
\begin{eqnarray}
g_b \equiv - [g] + [c]\, g'(c_{\infty}).
\label{3.19}
\end{eqnarray}
Recalling that $[g]$ and $[c]$ are functions of
radius $a$ as determined by thermodynamic jump
conditions, it is evident that the chemical
driving force (\ref{3.19}) is generally a
function of cluster radius, and {\em not a
constant} as in conventional nucleus energetics.

We can now write Eq.\ (\ref{3.6}) for $\dot{a}$ in terms of $G_a
= - \{ g'(c_{\infty}) - g'(c_+)\}\, 4\pi a^{2}[c]$ given by
(\ref{3.16}). The result is
\begin{eqnarray}
\dot{a} = - {\int_{c_{+}}^{c_{\infty}} D(c)\, dc\over
g'(c_{\infty}) - g'(c_+)}\, {G_{a}\over 4\pi [c]^{2} a^{3} }.
\label{3.16a}
\end{eqnarray}

\subsection{Energetics at small supersaturation}
In the small supersaturation limit, $\int_{c_{+}}^{c_{\infty}}
D(c)\, dc \sim D(c_1)\, (c_{+}-c_{\infty}) = D_1\, (\gamma_{+}-
\gamma_{\infty})$, $g'(c_{\infty}) - g'(c_+)\sim g''_1\, (
\gamma_{+} - \gamma_{\infty})$ and (\ref{3.16a}) reduces to
\begin{eqnarray}
\dot{a} = - {G_{a} D_1\over 4\pi g''_1 [c]^{2} a^{3} } .
\label{3.16b}
\end{eqnarray}
Inserting the approximate chemical driving force $g_b = [c]
g'(c_\infty) -[g] \sim g''_1 \gamma_{\infty}$ from (\ref{3.19})
into the free energy (\ref{3.17a}), we obtain 
\begin{eqnarray}
G \sim 4\pi\sigma a^2 - [c]\, g''_1\,
\gamma_{\infty} \, \left({4\pi\over 3}\, a^3
\right)\,.   \label{3.21}
\end{eqnarray}
Notice that the asymptotic chemical driving force, 
\begin{eqnarray}
g_b \sim [c]\, g''_1\,\gamma_{\infty},
\label{3.22a}
\end{eqnarray}
is in fact a constant independent of radius $a$, as in
the conventional wisdom. An alternative expression
for $g_b$ is useful: In certain experiments
surface tension $\sigma$ and critical radius
$a_c$ are measured observables, so it is
convenient to represent $g_b$ in terms of
$\sigma$ and $a_c$, 
\begin{eqnarray}
g_b = {2\sigma\over a_{c}}\,.  \label{3.22b}
\end{eqnarray}
This formula follows from the condition  $G'(a_c)
= 0$. The natural unit of free energy is $\sigma
l^2$. This is entered into the last column of the
scaling table. Then the dimensionless version of the
free energy formula (\ref{3.21}) is 
\begin{eqnarray}
G = 4\pi\, a^2 - \gamma_{\infty}\, {4\pi\over
3}\, a^3 .   \label{3.23}
\end{eqnarray}

\subsection{Identification of rate constant in BD kinetics}
We can now compare Eq.\ (\ref{dot-a}) for the growth of a
(large) cluster radius in BD kinetics with the corresponding
equations (\ref{3.16a}) and (\ref{3.16b}) obtained from our
macroscopic description. We find 
\begin{eqnarray}
k_d &=& {\tau\, \int_{c_{+}}^{c_{\infty}} D(c) dc\over
g'(c_{\infty}) - g'(c_{+})}\, \left( {c_{2}\over
[c]}\right)^{2}\, 4\pi a\label{5.15a}\\ 
& \sim & {D_{1}\tau\over g''_{1}}\, \left( {c_{2}\over
[c]}\right)^{2}\, 4\pi a ,   \label{5.15}
\end{eqnarray}
in the small supersaturation limit.

With this determination of $k_d$, the continuum
limit equation (\ref{5.12}) of cluster kinetics
is completely specified. It is easy to check that
the corresponding microscopic rate constant
determined by Penrose et al \cite{pen83,pen84}
is also proportional to $a$ for large clusters.  
One might wonder about the micromolecular basis of
(\ref{5.15}). While that requires further work,
here is a curious observation which might become
relevant to this question: Recall the {\em
mobility} $\delta$ defined in the formulation of
the macroscopic transport theory of Section
\ref{secMkinetics}. It is related to the diffusion
$D$ by (\ref{2.3a}), 
$$ D = \delta\, (1-\nu c)\, g''.$$
Hence the ratio $D_1/g''_1$ in (\ref{5.15}) is
given by
\begin{eqnarray}
{D_{1}\over g''_{1}} = \delta_1\, (1-\nu c_1).   
\label{5.16}
\end{eqnarray}
In the next Section, we show that the volume fraction $\nu
c_1$ of precipitate in matrix phase is small, $\nu c_1
\sim (0.04521$ nm$^3)\, (2.24$ nm$^{-3})\approx
0.10$, for coarsening experimental data in binary alloys
\cite{XH}. Then $D_1/g''_1\approx
\delta_1$ and formula (\ref{5.15}) for $k_d$ reduces to
\begin{eqnarray}
k_d \approx \delta_{1}\tau\left({c_{2}\over
[c]}\right)^{2}\, 4\pi a .  \nonumber
\end{eqnarray}

Let us notice that this rate constant $k_d$ is linear in the
cluster radius so that it scales as $n^{{1\over 3}}$ with
cluster size. This contrasts with the usual Turnbull-Fisher
rate constant that scales as $n^{{2\over 3}}$ \cite{reg01}, but
it agrees with the microscopic considerations of Penrose et al
\cite{pen83,pen84}. The scaling $n^{{1\over 3}}$ has been
shown to yield the Lifshitz-Slyozov distribution function for
cluster radii \cite{juanjo}. The latter is a roughly adequate
description of coarsening \cite{XH,gas01}. 

\section{Material and energy parameters of
kinetic theory determined from Xiao-Haasen data}
\label{sec-XH}

\subsection{Small supersaturation}
The nucleation in Xiao-Haasen's (XH) paper \cite{XH} takes
place under low supersaturation: The initial
sample has uniform composition, with mole
fraction $\chi\equiv 0.12$, or 12 \% of Al in a
Ni matrix. {\em Equilibrium} mole fraction of Al
in matrix phase at annealing temperature of 773 K
is $\chi_1 = 0.101$, or 10.1 \%. Equilibrium mole
fraction of Al in precipitate phase is $\chi_2 =
0.230$, or 23 \%. Hence, {\em supersaturation} as
a function of equilibrium concentration has
initial value
$${\chi-\chi_{1}\over \chi_{1}} \approx {0.120
-0.101\over 0.101} \approx 0.19.$$
Mole fractions are converted into number
densities: XH report a molar volume of
precipitate phase $V_m \approx 27.16\times
10^{-6}$ m$^3$. Conversion to an atomic volume by
Avogadro's number gives
$$ \nu_m\equiv {V_{m}\over N_{A}}\approx
4.51\times 10^{-29} \,\mbox{m}^3$$
or $\nu_m\approx 0.0451$ nm$^3$. 

XH also report a lattice constant of $a\approx
0.356$ nm for the precipitate phase, and atomic
volume corresponding to this lattice constant is
$\nu_m = a^3 \approx 0.0451$ nm$^3$. It is clear
that the molar volume $V_m$ was derived from the
lattice constant. A lattice constant for the
matrix phase is not reported explicitly, so it is
presumably close to the value $a\approx 0.356$ nm
of the precipitate phase. It seems there is an
implicit assumption: Local structure of alloy in
a lattice, with sites that can be occupied by Al
or Ni atoms. In this case, atomic volumes of Al
and Ni are de-facto the same, i.e., 
$$
\nu_m =\overline{\nu}_m \approx
0.0451\,Ê\mbox{nm}^3 .$$ 
Now number densities of Al and Ni easily follow.
For instance, $c_1$, the equilibrium number
density of Al in matrix phase is 
$$c_1 ={\chi_{1}\over \nu} \approx {0.101\over
0.0451}\,\mbox{nm}^{-3}\approx 2.24\,
\mbox{nm}^{-3} .$$
Table 1 gives initial concentration of
Al in matrix phase, and equilibrium
concentrations $c_1$ and $c_2$ of Al in matrix
and precipitate phases. 
\bigskip

\begin{tabular}{||c|c|c||cr||}    
\hline $c$& $c_1$& $c_2$\\
\hline
2.66 & 2.24 & 5.10   \\ \hline
\end{tabular}
\\

\noindent {\bf Table 1}: Number densities
(nm$^{-3}$).

\subsection{Nucleation energetics}
In XH, the nucleus energy takes the classic form 
\begin{eqnarray}
G = 4\pi\sigma a^2 -g_b\,\left( {4\pi\over 3}\,
a^{3}\right)\,.      \label{4.1}
\end{eqnarray}
Here $\sigma$ is surface tension. A value $\sigma
\approx 0.014$ J m$^{-2} = 1.4\times 10^{-20}$ J
nm$^{-2}$ is deduced from interpreting coarsening
data with the Lifshitz-Slyozov (LS) theory. XH deals
with chemical driving force $g_b$ in two ways: 

\noindent (i) {\em ``Experimental''}. The
distribution of nuclei in the space of their
radii goes through a transient phase with two
peaks, separated by a local minimum at about 1.2
nm. XH conjecture that the initial radius $a_c$
is in fact this 1.2 nm. Estimate of $g_b$ now
follows from (\ref{3.22a}), 
$$ g_b \sim {2\sigma\over a_{c}}\approx 2.33
\times 10^{-20}\, \mbox{J\, nm}^{-3} .$$

Given an experimental estimate of the critical
radius, $a_c \approx 1.2$ nm at outset, one can
estimate the number of atoms in the critical
nucleus, both Al and entrained Ni:
$$ {1\over \nu}\, {4\pi\over 3}\, a_c^3 \approx
160.$$
Of these, 23\% are Al, so there are 
$$ n_c= 0.23\times 160 = 37$$
Al atoms in the critical nucleus. It seems that
the critical nucleus is ``just big enough'' so
energetics based on continuum theory applies. 

\noindent (ii) {\em ``Theory''}. In standard
theories, $g_b$ is computed from both
thermodynamic properties of precipitate and bulk
phases. As such, it comes out as a constant
independent of nucleus radius $a$. These
derivations do not face up to the fine points of
the real situation, summarized in the formula
(\ref{3.19}) for $g_b$. So our approach is to
stick with the determination of $g_b$ based on
$\sigma$ and $a_c$, 
$$g_b \approx {2\sigma\over a_{c}}\approx 2.33
\times 10^{-20}\,\mbox{J\, nm}^{-3},$$ 
and then see what can be said about $g(c)$. In
(\ref{3.22a}) one can determine $g''_1$ because
all the other quantities are known. In fact, one
gets 
$$g''_1 = {g_{b}\over [c]\gamma_{\infty}} \approx
{2.33\times 10^{-20} \,\mbox{J\, nm}^{-3}\over
(5.10 - 2.24)\, \mbox{nm}^{-3}\, (2.66 - 2.24)\,
\mbox{nm}^{-3} } \approx 1.94\times 10^{-20} \,
\mbox{J\, nm}^{-3} .
$$
The annealing temperature of 773 K defines a
basic unit of energy, 
$$\tau = (773\,\mbox{K})\, ( 1.38\times 10^{-23}
\, \mbox{J/K}) \approx 1.07\times 10^{-20}\,
\mbox{J}.
$$
One now has
$${g''_{1}\over \tau}Ê\approx 1.81\,\mbox{nm}^3.
$$
This is just one number imposed upon free energy
function $g(c)$ by the XH data, but it is
sufficient to establish the \\
\\
{\em Nonideal character of Al solution in matrix
phase.}\\

Suppose the solution is ideal. Then the chemical
potential of an Al particle in matrix phase is
given by 
\begin{eqnarray}
\mu(c) = \mu_1 + \tau\ln {c\over c_{1}} ,
\label{4.2}
\end{eqnarray}
where $\mu_1$ is the chemical potential when $c=
c_1$ is the planar solvability. Now the relation
between $\mu(c)$ and $g(c)$ is given by
(\ref{1.2b}), which is repeated here for easy
reference, 
\begin{eqnarray}
\mu(c) = g'(c) + \nu\, (g - c\, g'). \label{4.3}
\end{eqnarray}
>From this equation is evident that $g''(c)$ gives
information about $\mu'(c)$. In fact,
differentiation of (\ref{4.3}) yields 
\begin{eqnarray}
\mu'(c) = (1- \nu c)\, g''(c)\Longrightarrow
\mu'_1 = (1-\nu c_1)\, g''_1 .\label{4.4}
\end{eqnarray}
Numerical value of $\mu'_1/\tau$ based upon
previous value of $g''_1$ turns out to be 
$${\mu'_{1}\over\tau} \approx \{ 1- (0.0451 \,
\mbox{nm}^3)\, (2.24 \,\mbox{nm}^{-3})\}\, (1.81\,
\mbox{nm}^3)\approx 1.63\, \mbox{nm}^3 .
$$
If the ideal solution formula (\ref{4.2}) were
correct, one would get 
$${\mu'_{1}\over\tau} = {1\over c_{1}} \approx {1
\over 2.24 \,\mbox{nm}^{-3} }\approx 0.45\,
\mbox{nm}^3 ,
$$
which is 1/274 of value that follows from XH
parameters. That the solution of Al in Ni phase
is not ideal was already known to XH. They in
fact considered that our chemical driving force
$g_b$ is sum of two terms: (i) a chemical driving
force estimated from the activity of Al component
at the concentrations $\chi$ and $\chi_1$, and
(ii) and the elastic strain energy per unit
volume. With the corresponding expressions, they
obtained a value for the critical radius, $a_c
\approx 1.7$ nm, which is not too far from the
experimental value, $a_c \approx 1.2$ nm
\cite{XH}. 

The experimentally derived values of surface
tension $\sigma$ and chemical driving force
$g_b$ in (\ref{4.1}) set important parameters for
macroscopic nucleation theory, namely: Energy
barrier for nucleation, and typical free energy
cost to add one Al particle to a nucleus. 

{\em Energy barrier} is given by 
$$G_{\mbox{nuc}} = G(a_c) = {4\pi\over
3}\,\sigma a_c^2 \approx {4\pi\over 3}\,
(1.4\times 10^{-20}\,\mbox{J\, nm}^{-2})\, (1.2\,
\mbox{nm})^2 \approx 8.4\times 10^{-20}\,\mbox{J}
.  $$
Energy barrier in units of thermal energy is 
$${G_{\mbox{nuc}} \over\tau} \approx 7.8 .
$$
A reasonable looking number. Notice that
exponential 
$$ e^{ - {G_{\mbox{nuc}} \over\tau} } \approx
3.7\times 10^{-4}, $$
which appears in the nucleation rate is not too
small. This makes anthropomorphic sense: In XH
experiment, nucleation kinetics unfolds in hours
and days time scales, not unduly taxing to
humans. Evidently, the annealing temperature is
tuned so as to achieve a ``reasonable''  nucleation
rate. 
\bigskip

{\em Free energy cost to add one particle.}

The number $n$ of Al particles in nucleus is
related to radius $a$ by 
\begin{eqnarray}
n = {4 \pi\over 3}\, c_2 a^3 \Longrightarrow a =
\left({3n\over 4\pi c_{2}}\right)^{{1\over 3}} .
\label{4.5}
\end{eqnarray}
Substituting (\ref{4.5}) for $a$ in (\ref{4.1})
gives nucleus energy as a function of $n$,
\begin{eqnarray}
G_n = (36 \pi)^{{1\over 3}}\,\sigma c_2^{-{2\over
3}} \, n^{ {2\over 3}} - g_b\, c_{2}^{-1}\, n.
\label{4.6}
\end{eqnarray}
Free energy cost to add one particle to nucleus
of $n$ particles, in units of thermal energy
$\tau$, is
$${\mu_{n}\over\tau} \equiv {G_{n+1} - G_{n}
\over\tau}  = (36 \pi)^{{1\over 3}}\,\sigma 
c_2^{-{2\over 3}}\tau^{-1} \,\{ (n+1)^{ {2\over 3}}
-  n^{ {2\over 3}}\} - g_b\, c_{2}^{-1}\,\tau^{-1}.
$$
Since this formula is based on continuum theory,
its validity requires $n\gg 1$, in which case it
reduces to 
\begin{eqnarray}
{\mu_{n}\over\tau} \sim {2\over 3} (36
\pi)^{{1\over 3}}\,\sigma c_2^{-{2\over 3}}\,
\tau^{-1} n^{ -{1\over 3}} - g_b\, c_{2}^{-1}\,
\tau^{-1}.    \label{4.7}
\end{eqnarray}
Substituting XH parameter values in RHS,
\begin{eqnarray}
{\mu_{n}\over\tau} \sim 1.43\, n^{ -{1\over 3}} -
0.42.    \label{4.8}
\end{eqnarray}
In the limit $n\to\infty$, we get $|\mu_n/\tau|
\sim 0.42$. While less than 1, one would not call
this value ``small compared to one''. For $n=37$,
corresponding to a critical nucleus, of course
one gets $\mu_n/\tau=0$. Hence there will be a
range of $n$ about $n= n_c = 37$ in which $|\mu_n
/\tau|\ll 1$. In particular, $|\mu_n/\tau| < 0.2$
in the rather generous interval $12<n<300$. In
this range of $n$, asymptotic reduction of
discrete kinetic  models such as Becker-D\"oring
to a Smoluchowski partial differential equation (PDE) should be
reasonable.

\section{Discussion}
\label{sec-BDasymptotics}
In coarsening experiments, one starts from a
situation of equilibrium at high temperature in
which most clusters are monomers. Then the
temperature is lowered to a value below the
critical temperature, and kept there. Clusters are
nucleated and grow, supersaturation changes with
time so that nucleation of new clusters becomes
unlikely, and the coarsening of clusters proceeds.
As explained by Penrose et al \cite{pen83,pen84},
this process is reasonably well described by the
Becker-D\"oring model (better than by the
Lifshitz-Slyozov distribution function), provided
the volume fraction of precipitate is small. Let us describe
the nucleation and coarsening processes in typical experiments
such as XH and which parts thereof are mathematically
understood. 

The nucleation process described by the BD equations starts at
$t=0$ with some initial value of $\rho_1= c_{\infty}$ and no
supercritical clusters. According to the XH data, the energy
barrier corresponding to the initial value of the critical
nucleus is relatively high, $G_{\mbox{nuc}}/\tau
\approx 7.8$, so that we may consider the clusters
below critical size ($n<n_c$) to be in a
quasistationary state. The flux across the energy
barrier is then uniform and it supplies the source
for coarsening of clusters larger than the
critical size. As explained in Section
\ref{sec-XH}, there is a range of sizes (about the
critical size) for which we may approximate the
discrete BD kinetics by a continuum Smoluchowski
equation for the distribution function
$\rho$. The latter will describe the coarsening
process and it should be approximately solved with a
boundary condition obtained by matching to the
solution of the BD equations for $n<n_c$. 
For $t>0$, supercritical clusters are
created at the rate $j$ per unit volume given in Eq.\
(\ref{6.6}) below, and $\rho_1$ starts to decrease. A small
change of $\rho_1$, $O(1/n_c)=O(\epsilon^3 l^3/c_2)$, induces
an $O(1)$ relative change of $j$. 
%Hence we adopt the
%nondimensionalization suggested by the Scaling
%Table of Section \ref{secMkinetics}, so that the
%continuum limit of BD is the problem (\ref{5.17})
%to (\ref{5.19}). We would like to describe the
There is a transient situation during which 
$\rho$ becomes a bimodal distribution function with peaks at sub
and supercritical sizes. As time evolves, the
supercritical peak increases at the expense of
the subcritical peak, which disappears given
enough time. Then the resulting unimodal
distribution evolves toward a function with the LS
scaling. 

Currently it is known that the LS distribution function
\cite{LS} is a solution of the Smoluchowski equation for a very
special boundary condition at small cluster size \cite{juanjo}.
Although the stability properties of the LS distribution
function are not completely elucidated, it seems clear that the
Smoluchowski equation may have other stable solutions that may
match the quasistationary distribution at small cluster sizes.
The appropriate solution of the Smoluchowski equation should
then describe the transient stage of coarsening. As the time
advances, the peak of the distribution function at subcritical
sizes decreases and disappears while the peak at supercritical
sizes takes over. The latter should have the LS scaling to
explain experimental \cite{XH} and numerical data
\cite{pen83,pen84}. To carry out an asymptotic analysis of
nucleation and coarsening providing the same qualitative
description sketched here is a challenging future task.

 \section{Acknowledgements}
 The present work was financed through the Spanish
DGES grant PB98-0142-C04-01 and carried out during J.\ Neu's
sabbatical stay at Universidad Carlos III supported by the
Spanish Ministry of Education.

%\appendix
\section{Appendix: BD kinetics for $n<n_c$}
The quasistationary state is a solution of the BD
equations characterized by uniform flux,
\begin{eqnarray}
j_{n} = k_{d,n+1}\,\left\{ e^{-{G_{n+1} -
G_{n}\over\tau}}\,\rho_{n} - \rho_{n+1}
\right\} \equiv j,   \label{6.1}
\end{eqnarray}
for $n<n_c = c_2 4\pi a_c^3/3$. At high
temperature, before the experiment starts, we have
the following equilibrium solution
\begin{eqnarray}
\rho_{eq,n} = {c\, e^{-{G_{n}\over\tau}}\over
\sum_{l=1}^{\infty} l\, e^{-{G_{l}\over\tau}}}
\sim c\, e^{-{G_{n}-G_{1}\over\tau}} . \label{6.2}
\end{eqnarray}
To write the above approximation, we have assumed
that $G_2/\tau\gg G_1/\tau\gg 1$ and that
$G_n$ increases with $n$. After nucleation and
coarsening start, we shall assume that $\rho_n$ is
close to its equilibrium value [given by the
approximate expression (\ref{6.2}) at the correct
temperature $\tau$], as $n\ll n_c$. Thus $\rho_n =
O(c\, e^{-(G_{n}-G_{1})/\tau})$ if $n<n_c$, and
much smaller than this order if $n>n_c$. This
means that $e^{{G_{n}\over\tau}}\,\rho_{n}/c =
O(e^{G_{1}/\tau})$ if $n <n_c$, and that
$e^{{G_{n}\over\tau}}\,\rho_{n}/c = o(e^{G_{1}/
\tau})$ if $n\gg n_c$. 

Equation (\ref{6.1}) can be written as
$$ e^{{G_{n+1} \over\tau}}\,\rho_{n+1} - 
e^{{G_{n}\over\tau}}\,\rho_{n} = -
{j\over k_{d,n+1}}\, e^{{G_{n+1}\over\tau}}\,,
$$
and therefore easily integrated under the
condition $e^{{G_{n}\over\tau}}\,\rho_{n} \to 0$
as $n\to\infty$:
\begin{eqnarray}
e^{{G_{n}\over\tau}}\,\rho_{n} = j\,
\sum_{l=n}^\infty { e^{{G_{l+1}\over
\tau}}\over k_{d,l+1}}\,.   \label{6.3}
\end{eqnarray}
The terms in this sum are largest for $l\sim n_c$,
at which $G_l$ is maximum. For such integers, the
continuum approximation holds, and we can write
\begin{eqnarray}
{ e^{{G_{l+1}\over\tau}}\over k_{d,l+1}}\sim
{e^{{G_{n_{c}}\over\tau}}\over k_{d,n_{c}}}
\sum_{l=n}^\infty e^{-{4\pi\sigma\over\tau}\,
(a_{l}-a_{c})^{2}}.    \label{6.4}
\end{eqnarray}
We have used $G_{l+1} - G_{n_{c}}\sim - 4\pi\sigma 
(a_l-a_c)^2$, for $l+1$ close to $n_c$. We now
approximate $a_l = a_c + x\sqrt{\sigma/\tau}$ in
(\ref{6.3}), and $1\sim 4\pi c_2 a_c^2 da_l = 4\pi
c_2 a_c^2 \sqrt{\tau/\sigma}\, dx$, so that
(\ref{6.3}) becomes 
\begin{eqnarray}
e^{{G_{n}\over\tau}}\,\rho_{n} \sim {j\over
D_{1}}\, {[c]\over c_{2}}\,{\sqrt{ {\sigma\over
\tau}}\over \gamma_{\infty}}\, e^{{G_{n_{c}}\over
\tau}}\, 2\, \int_{\sqrt{{\sigma\over\tau}}\,
(a-a_{c})}^\infty  e^{-4\pi x^{2}}\, dx.  
\label{6.5}
\end{eqnarray}
The equilibrium solution of the BD equations is 
(\ref{6.2}). If we impose that $\rho_n \sim
\rho_{eq,n}$ as $n\ll n_c$, (\ref{6.2}) and
(\ref{6.5}) yield 
\begin{eqnarray}
j\sim \sqrt{{\tau\over\sigma}}\, {c D_{1}
\gamma_{\infty} c_{2}\over[c]}\,
e^{-{G_{n_{c}}-G_{1}\over\tau}} \,. 
\label{6.6}
\end{eqnarray}
This constant flux is exponentially small because$(G_{n_{c}}
-G_1)/\tau \sim G(a_c)/\tau \gg 1$. Notice that it is also
proportional to the supersaturation $\gamma_\infty$. It is
clear that a small change in the supersaturation,
$\delta\gamma = O(\gamma_{\infty}/n_c)$, produces
an $O(1)$ change in $n_c$ and in $G_{n_{c}}$,
$\delta n_c = - 3 n_c \delta\gamma/
\gamma_{\infty}$ and $\delta G_{n_{c}} = - 2
G_{n_{c}} \delta\gamma/\gamma_{\infty}$, and hence
a significant relative change of $j$ in
(\ref{6.6}):
\begin{eqnarray}
{\delta j\over j}Ê\sim \exp\left({g''_{1} [c]
n_{c}\over\tau}\,\delta\gamma\right)\,.
\label{6.6b}
\end{eqnarray}

Notice that, in the continuum limit, the flux
$j_n$ becomes 
\begin{eqnarray}
j_n &\sim & -{D_{1}\tau\over g''_{1}[c]^{2}}\, {e^{-
{G(a)\over\tau}}\over 4\pi a}\, {\partial\over \partial a}
\left( e^{{G(a)\over\tau}}\, {\rho\over a^{2}}\right)\nonumber\\
&=& {D_{1}\tau\over g''_{1}[c]^{2}}\, {1\over 4\pi a}\,
\left( {\rho\over\tau a^{2}}\, {\partial G\over \partial a}
+ {\partial\over
\partial a}\left( {\rho\over a^{2}}\right)\right)\,. 
\label{6.6d}
\end{eqnarray}
The relation between drift and diffusion coefficients here is
$G_a/\tau$ in agreement with the formulas provided by
Nonequilibrium Thermodynamics; see Ref.\ \cite{reg01}.

%\end{multicols}
\end{document}